\documentclass[aps,amssymb,amsmath, twocolumn, prd, superscriptaddress]{revtex4}
\usepackage{graphicx}
\usepackage{epsfig}
\usepackage{amsmath}
\usepackage{bm}
\usepackage{bbm}
\usepackage{ulem}
\usepackage[colorlinks, citecolor = magenta]{hyperref}
\usepackage{hyperref}
\usepackage{color}
\usepackage{soul}
\usepackage{centernot}
\usepackage[up]{subfigure}
\usepackage{footnote}

\usepackage{epstopdf}

\usepackage{lipsum}

\newcommand{\be}{\begin{equation}}
\newcommand{\ee}{\end{equation}}
\newcommand{\bc}{\begin{center}}
\newcommand{\ec}{\end{center}}
\newcommand{\bea}{\begin{eqnarray}}
\newcommand{\eea}{\end{eqnarray}}
\newcommand{\ba}{\begin{array}}
\newcommand{\ea}{\end{array}}

\def\bra#1{\mathinner{\langle{#1}|}}
\def\ket#1{\mathinner{|{#1}\rangle}}
\def\braket#1{\mathinner{\langle{#1}\rangle}}

\usepackage{booktabs}

\usepackage{amsthm}
\begin{document}
\title{Neutrino oscillations in discrete-time quantum walk framework}
\author{Arindam Mallick}
\email{marindam@imsc.res.in}
\author{Sanjoy Mandal}
\email{smandal@imsc.res.in}
\author{C. M. Chandrashekar}
\email{chandru@imsc.res.in}
\affiliation{The Institute of Mathematical Sciences, C. I. T. Campus, Taramani, Chennai 600113, India}
\affiliation{Homi Bhabha National Institute, Training School Complex, Anushakti Nagar, Mumbai 400094,  India}
\begin{abstract}
 Here we present neutrino oscillation in the frame-work of quantum walks. Starting from a one spatial dimensional 
 discrete-time quantum walk we present a scheme of evolutions that will simulate neutrino oscillation. 
 The set of quantum walk parameters which is required to reproduce the  oscillation probability profile obtained in both, 
 long range and short range neutrino experiment is explicitly presented. Our scheme to simulate three-generation neutrino oscillation 
 from quantum walk evolution operators can be physically realized in any low energy experimental set-up with 
 access to control a single six-level system, a multiparticle three-qubit or a qubit$-$qutrit system. 
 We also present the entanglement between spins and position space, during neutrino propagation 
 that will quantify the  wave function delocalization around instantaneous average position of the neutrino.
 This work will contribute towards understanding neutrino oscillation in the framework of the quantum information perspective. 
 \end{abstract} 
  
\maketitle
\section{Introduction} 
\label{intro}

Neutrino oscillation is a well established phenomenon explained by quantum field theory (QFT). 
Pauli first proposed the neutrino to explain the continuous spectrum of electron in beta decay~\cite{pauli}.
In the standard model (SM) description neutrinos are massless and a very weakly interacting particles.
However, in order to give a correct interpretation to  the experimental results it was established
that the neutrinos are massive and leptons  mix~\cite{SKM Collaboration,SNO Collaboration,KamLAND Collaboration}. 
Neutrino oscillation, which implies neutrino can change from one flavor to another, is 
also a consequence of the neutrino masses and lepton mixing~\cite{Pontecorvo}. 
Hence, neutrino oscillations indicate an incompleteness of the SM and opens a window for physics beyond SM. 

The last few years have seen an increasing interest by the physics community in reconstructing and 
understanding physics from a quantum information perspective, that is to say, 
``It from qubit''~\cite{itbit, feyn, cirac}. This approach will give us access to understanding various natural
physical processes in the form of quantum information processing.
This will also facilitate us to simulate inaccessible and experimentally demanding high energy phenomena
in low energy quantum bit (qubit) systems. Parameter tuneability in protocols that can simulate real effects allows access to different physical regimes  which 
are not accessible in real particle physics experiments.
Thus, we can anticipate a significant contribution to our understanding of physics beyond known standard theories via quantum simulations.  
 
Any standard quantum information processing protocols on a basic unit of quantum information, 
that is, the qubit, can be described using three steps : (a) initial state preparation, (b) evolution operations, and (c) measurements. 
In this work, starting from initial state preparation of qubit we will present 
a scheme of evolution that can simulate the three-flavor neutrino oscillation. 
To simulate neutrino oscillations where the dynamics of each flavor is defined by the Dirac equation, 
we will use discrete-time quantum walk (DTQW) evolution operators. 
The DTQW, defined as the evolution of wave packet in a superposition of position space can also be 
viewed as a physical dynamics of information flow \cite{dwalkin1, dwalkin2} which can be engineered 
to simulate various quantum phenomena, for example, like Anderson localization~\cite{anderson}, 
the Dirac equation~\cite{Molfetta,cm,Marc,subha}, and topologically bound states~\cite{ topological}. 
Recent results have shown that one-dimensional DTQWs
produce the free Dirac equation in 
the small mass and momentum limit~\cite{cm}. The neutrino mass eigenstates being the solution 
of the free Dirac Hamiltonian was the main motivation for us to connect the simulation of the free Dirac equation 
and neutrino oscillation with the DTQW. This will also allow us to understand neutrino oscillation 
in the framework of DTQW whose dynamics is discrete- in both, space and time.
The description of the dynamics in the form of a unitary operation for each discrete 
time step will help us to address the quantum correlations between the position space 
and spin degree of freedom as a function of time. This discrete approach will also lead towards
simulating various high energy phenomena and addressing the dynamics of quantum correlations between different
possible combination of Hilbert spaces involved in the dynamics.

To simulate oscillations between three neutrino flavors we present DTQW on a system with six internal 
degrees of freedom which physically can be realized using a single six-dimensional system or a three-qubit system or a qubit$-$qutrit system.
With DTQW being experimentally implemented in various physical systems~\cite{tion, catom, nmr, photon},
simulation of the long and short range neutrino oscillations on different low energy physical systems will also easily be 
realizable as a function of the number of walk steps.
By preparing different initial states, different types of neutrino oscillations can be simulated
in a simple table-top experimental set-up which are not straightforward in real world neutrino oscillation experimental set-ups.

For the DTQW parameters which will simulate neutrino oscillation, we will calculate the entropy
of the density matrix as a function of DTQW steps. This entropy which captures information content 
of the evolution of the neutrino density matrix can be effectively 
used to quantify the wave function delocalization in position space.
We will also  calculate the correlation entropy between the position space and particular neutrino flavor 
as measure of possible information extractable about the whole state of the neutrino wave function
when we detect spin part of that particular flavor state.  
This simulation and information from entropy will contribute towards understanding the role 
of quantum correlations in neutrino oscillations reported recently~\cite{alok, subhasish}. 
Exploring neutrino physics in general from quantum information perspective will easily be accessible.
This could further lead to a way to use quantum simulations and quantum information to study physics beyond SM 
and understand the quantum mechanical origin of the some of the interesting phenomena in nature.

The paper is organized as follows. In Sect. \ref{neuphy} we present a brief introduction of the neutrino oscillation theory. 
In Sect. \ref{DQW}, we introduce the DTQW evolution, we use for simulation of neutrino oscillation.
In Sect. \ref{mimic} we present the scheme for simulation of the three-flavor neutrino oscillation using one spatial dimensional DTQW.
In Sect. \ref{nsimulate}, we numerically simulate these oscillations and present 
the DTQW evolution parameters that recover the short and long range neutrino oscillations.
In Sect. \ref{entanglement} we present entanglement between position space and internal degrees of freedom of neutrino during propagation.  
Finally we will end with concluding remarks in Sect. \ref{conclu}.


\section{Physics of the neutrino oscillation}
\label{neuphy}


Here, we give a brief discussion of the theory of neutrino oscillations~\cite{mohapatra,boris}.
So far, experimentally three flavors of the neutrinos,
$\nu_{e}$, $\nu_{\mu},$ and $\nu_{\tau}$, have been detected.
The neutrino of a given flavor are defined by the leptonic $W$-boson decay. 
That is,  the $W$-boson decays to a charged lepton (e, $\mu$, or $\tau$) and a neutrino. 
We will define a neutrino as $\nu_{e}$, $\nu_{\mu}$, or $\nu_{\tau}$ when the corresponding charged leptons are e, $\mu$, or $\tau$.
Studies have reported that neutrinos have masses and lepton mixing means that
there is some spectrum of the neutrino mass eigenstates $ \ket{\nu_{i} }$.
Using this we can write the neutrinos of definite flavor as a quantum superposition
of the mass eigenstates~\footnote{For the case of ultra-relativistic 
neutrinos, we can derive this standard flavor state (Eq.~\ref{flavor state})
in a more rigorous way in the framework of quantum field theory~\cite{C Giunti}.}
\begin{equation}\label{flavor state}
 \ket{\nu_{\alpha}}=\sum_{j}U_{\alpha j}^{*}\ket{\nu_{j}}
\end{equation}
where $\alpha = e$, $\mu$, $\tau,$ and $ j = 1, 2, 3$. $ U_{\alpha j}^{*}$ is the complex conjugate of the $\alpha j$ th component 
of the matrix $U$. $ U $ is a $ 3 \times 3 $ unitary matrix and is referred to as the Maki$-$Nakagawa$-$Sakata (MNS) matrix, 
or as the Pontecorvo$-$Maki$-$Nakagaea$-$Sakata (PMNS) matrix~\cite{pmns}. The matrix $U$ and its decomposition can be written as
\begin{widetext}
\begin{align}\left[ 
  \begin{array}{ccc}
    U_{e1} & U_{e2} & U_{e3} \\
    U_{\mu 1} & U_{\mu 2} & U_{\mu 3}\\
    U_{\tau 1} & U_{\tau 2} & U_{\tau 3}
  \end{array}\right]= 
 \begin{bmatrix}
  1 & 0 & 0\\
  0 & c_{23} & s_{23}\\
  0 & -s_{23} & c_{23}
 \end{bmatrix}
 \begin{bmatrix}
  c_{13} & 0 & s_{13}e^{-i\delta}\\
  0 & 1 & 0\\
  -s_{13}e^{i\delta} & 0 & c_{13}
 \end{bmatrix}
 \begin{bmatrix}
  c_{12} & s_{12} & 0\\
  -s_{12} & c_{12} & 0\\
  0 & 0 & 1
\end{bmatrix}
\begin{bmatrix}
 e^{i\alpha_{1}/2} & 0 & 0\\
 0 & e^{i\alpha_{2}/2} & 0\\
 0 & 0 & 1
\end{bmatrix},
\end{align}
  \end{widetext}
where $c_{ij}\equiv\text{cos}\,\theta_{ij}$ and $s_{ij}\equiv\text{sin}\,\theta_{ij}$ with  $\theta_{ij}$ being the mixing angle, 
and $\alpha_{1}$, $\alpha_{2}$, and $\delta$ are CP-violating phases.
The state $ \ket{\nu_{j}} \in \mathcal{H}_\mathrm{spin} \otimes span\{ \ket{k_{j1}}, \ket{k_{j2}}, \ket{k_{j3}}\} $ 
is the mass eigenstate of the free Dirac Hamiltonian,
\begin{align}\label{hamil}
 H_j =  \vec{\xi}.~ \hat{\vec{ p}}_j~ c + \beta m_j c^2 
\end{align}
where $c$ is the velocity of light in free space, $m_j$ is the mass,
$ \hat{\vec{p}}_j$ is momentum operator corresponding to $j$ th particle,
with positive energy eigenvalue $E_j = \sqrt{|\vec{k}_j|^2 c^2 +  m_j^2 c^4}$.
Its propagation is described by the plane wave solution of the form
\begin{equation}
 \braket{ \vec{x} |\nu_{j}(t)}=e^{-\frac{i}{\hbar} \left(E_{j}t-\vec{k}_{j}.\vec{x}\right)}\braket{ \vec{x} | \nu_{j}(0)}
\end{equation}
where $t$ is the time of propagation, $\vec{k}_{j}$ is the three momenta, and $\vec{x}$ is the position of the particle in the mass
eigenstate from the source point. As the neutrino mass is very small, they are ultra-relativistic
particles ($|\vec{k}_{j}| \gg m_{j} c $) and we can approximate the energy as
\begin{equation}
\label{energy approximation}
 E_{j}=\sqrt{ |\vec{k}_{j}|^{2} c^2 + m_{j}^{2} c^4 }\simeq |\vec{k}_{j}| c +\frac{m_{j}^{2} c^3 }{2k_{j}}\approx E+\frac{m_{j}^{2} c^4 }{2E}
\end{equation}
where $E \approx |\vec{k}_j| c$, and the same for all $j$ (taken for simplicity).
Now, consider a neutrino beam $\nu_{\alpha}$ which is created in a charged current interaction.  
After time $t$, the evolved state is given by
\begin{equation}
\begin{split}
\braket{\vec{x} | \nu_{\alpha}(t)}&=\sum_{j}U_{\alpha j}^{*}e^{- \frac{i}{\hbar}\left(E_{j}t-\vec{k}_{j}.\vec{x}\right)}\braket{\vec{x} | \nu_{j}}\\
&=\sum_{\beta}\sum_{j}U_{\alpha j}^{*}e^{- \frac{i}{\hbar} \left(E_{j}t-\vec{k}_{j}.\vec{x}\right)}U_{\beta j}\braket{ \vec{x} | \nu_{\beta}}.
\end{split}
\end{equation}
For simplicity we will work with one-dimensional space, so our choice will be $ \vec{k}_j = (k , 0, 0),$ the same for all $j$.  
This means
\begin{align}
 \braket{k | \nu_{\alpha}(t)} = \sum_{\beta}\sum_{j}U_{\alpha j}^{*}e^{- \frac{i}{\hbar}  E_{j}t }U_{\beta j}\braket{ k | \nu_{\beta}}.
\end{align}

\begin{figure}
\subfigure[]{\includegraphics[width=8cm]{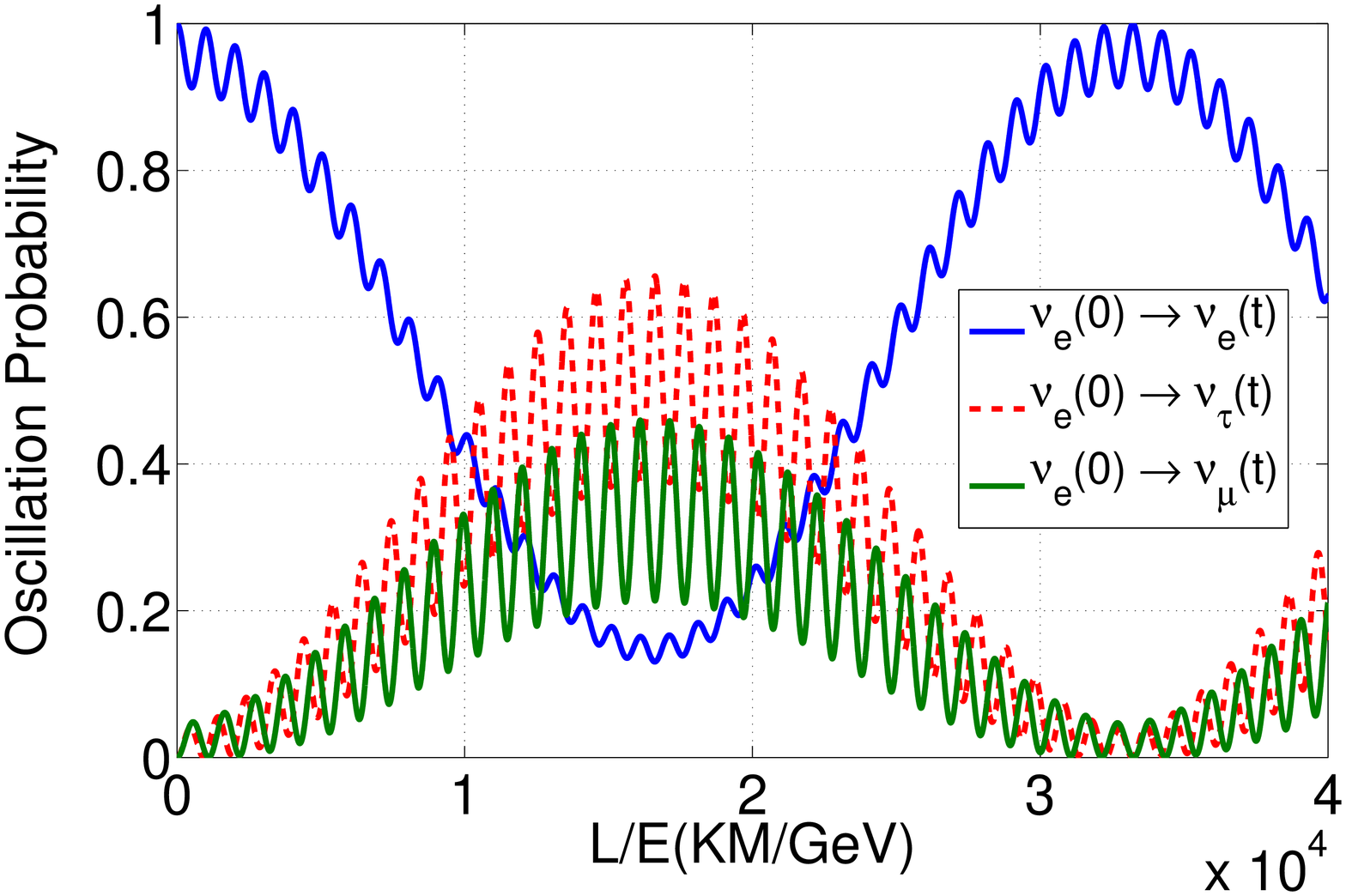}}
\subfigure[]{\includegraphics[width=8cm]{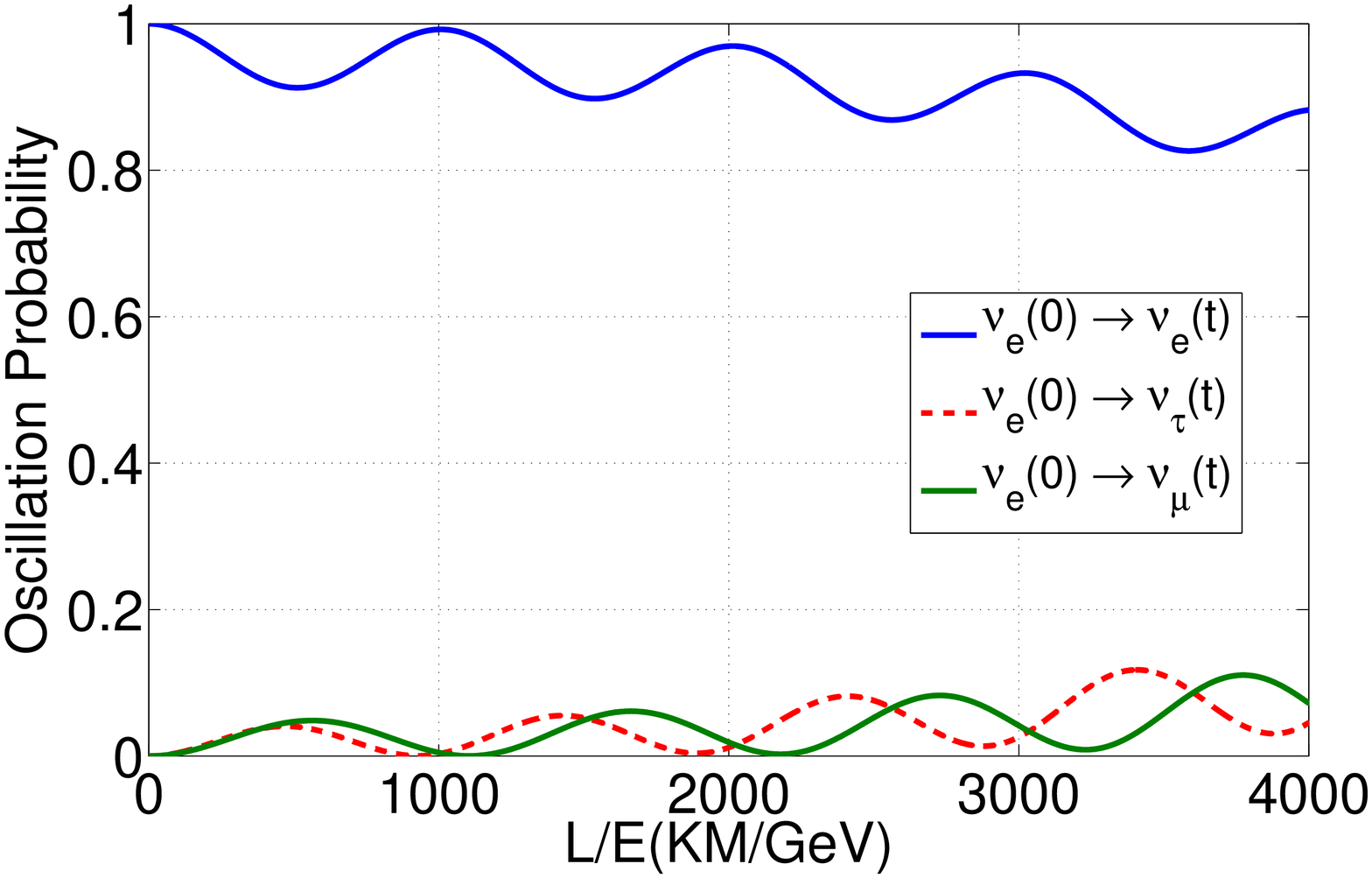}}
\caption{\small{Neutrino oscillation probabilities for an initial electron neutrino. 
Here, we show the oscillation probability $\nu_{e}(0)\rightarrow\nu_{e}(t)$ {\it (blue)},
$\nu_{e}(0)\rightarrow\nu_{\mu}(t)${\it (green)}, $\nu_{e}(0)\rightarrow\nu_{\tau}(t)$ {\it (red)}. 
The {\it left} figure {\bf (a)} gives long range neutrino oscillation and the {\it right} figure {\bf (b)} gives
short range neutrino oscillation.}}
\label{oscillation plot}
\end{figure}

As neutrinos are ultra-relativistic particles,
we can also replace $ c \times t $ by the traveled distance $x$. Using these assumptions and Eq.~(\ref{energy approximation}),
the amplitude of finding flavor state $\ket{\nu_{\beta}}$ in the original $\ket{\nu_{\alpha}}$ beam at time $t$, it is given by
\begin{equation}
\braket{\nu_{\beta}|k} \braket{k | \nu_{\alpha}(t)}=\sum_{j}U_{\alpha j}^{*}e^{-im_{j}^{2} c^3 \frac{L}{2 E \hbar}}U_{\beta j}
\end{equation}
where, $L \approx c t $ is the travelled distance. 
Squaring it we find the transition probability  $\nu_{\alpha}(t = 0)\rightarrow\nu_{\beta}(t)$ and is given by
\begin{widetext}\begin{align}\label{oscillation formula}
 P\left(\nu_{\alpha}\rightarrow\nu_{\beta}\right)&=\delta_{\alpha\beta}- 4\sum_{j>r}Re\left(U_{\alpha j}^{*}U_{\beta j}U_{\alpha r}U_{\beta r}^{*}\right)
 \sin^{2}\left(\Delta m_{jr}^{2}\frac{L c^3}{4E \hbar}\right)
  +2\sum_{j>r}Im\left(U_{\alpha j}^{*}U_{\beta j}U_{\alpha r}U_{\beta r}^{*}\right)\sin\left(\Delta m_{jr}^{2}\frac{L c^3}{2E \hbar}\right),
\end{align}\end{widetext}
where $\Delta m_{jr}^{2}\equiv m_{j}^{2}-m_{r}^{2}$. If there is no CP violation we can choose the mixing matrix $ U $ to be real.
This will ensure that we will not have an imaginary part in the oscillation formula.
Including the factors $\hbar$ and $c$, we can write the argument of 
the oscillatory quantity $\sin^{2}\left(\Delta m_{jr}^{2}\frac{L c^3 }{4E \hbar}\right)$
that appears in Eq.~(\ref{oscillation formula}) by 
\begin{equation}\label{phase}
 \Delta m_{jr}^{2}\frac{L c^3 }{4E \hbar}=1.27\Delta m_{jr}^{2}(eV^{2})\frac{L(Km)}{E(GeV)}.
\end{equation}
So for large $ L/E $, neutrino oscillation provides experimental access to very tiny neutrino masses. In Fig.~\ref{oscillation plot},
we show the probabilities for the initial flavor $\ket{\nu_{e}}$ to be in $\ket{\nu_{e}}$, $\ket{\nu_{\mu}}$, $\ket{\nu_{\tau}}$ after time $t$,
with Fig.~\ref{oscillation plot} a showing long range and Fig.~\ref{oscillation plot} b
showing short range oscillation. These probability transition 
oscillation plots are obtained assuming the Normal Ordered (NO) neutrino mass spectrum ($m_{3}>m_{2}>m_{1}$).
The oscillation parameters used here are given as follows~\cite{Global Analyses of Neutrino Oscillation Experiments}:
\begin{align}\label{massdif}
 \Delta m_{21}^{2} = 7.50\times 10^{-5} \,\text{eV}^{2},~\\
 \Delta m_{31}^{2} = 2.457\times 10^{-3} \,\text{eV}^{2},\\
 \Delta m_{32}^{2} = 2.382\times 10^{-3}\,\text{eV}^{2},\\
 E=1\,\text{GeV}.~~~~~~~~
\end{align}
As $\delta$ has not been determined by experiments, 
it can take a value anywhere between 0 to 2$\pi$ and for simplicity we have taken $\delta = 0$ for our oscillation plots. 

\section{ Discrete-time quantum walk}
\label{DQW}

The discrete-time quantum walk (DTQW) is a quantum analogue of the classical random walk which evolves the particle
with an internal degree of freedom in a superposition of position space. 
The Hilbert space on which the walk evolves is $\mathcal{H}=\mathcal{H}_{c} \otimes \mathcal{H}_p$ 
where $\mathcal{H}_{c}$ is spanned by the internal degrees of 
freedom of the particle which hereafter will be called a coin space and $\mathcal{H}_{p}$ is
spanned by the position degree of freedom. Each step of the DTQW 
evolution operator $W$ is a composition of the quantum coin operator $C$ and  a coin dependent position shift operator $S$,
\begin{align}
 W = S ( C \otimes I ). 
 \end{align}
The identity operator $I$ in $(C \otimes I)$ acts only on spatial degree of freedom and operator $C$ acts only on the coin space.
The operator $C$ evolves the particles basis states to the
superposition of the basis states and the operator $S$ shifts
the particle to the superposition of position states depending on the basis states of the particle. 
DTQW on a one-dimensional space is commonly defined on a particle with two internal 
degrees of freedom, $|\downarrow \rangle$ and $|\uparrow \rangle$.
Therefore, the coin operation $C$ can be any $2\times 2$ unitary operator and the shift operator
which shifts the state by a distance $a$ in position space will be of the form
 \begin{align}
 S =  \ket{\downarrow} \bra{\downarrow} \otimes T_{-} + \ket{\uparrow} \bra{\uparrow} \otimes T_{+}
\end{align}
where 
\begin{align}
T_- = \sum_{x \in  a\mathbb{Z}} \ket{x - a} \bra{x} \mbox{,~} T_+ = \sum_{x \in  a\mathbb{Z}} \ket{x + a} \bra{x}= T^{\dagger}_{-}.
\end{align}
The operator $T_-$ shifts the particle position to one step farther along the negative $x$-axis and $T_+$ 
shifts the particle position one step along the positive $x$-axis. 
This standard definition of 
DTQW evolves quadratically faster in position space when compared to the classical random walk. 
This description has been extended to systems with both higher spatial dimension 
and higher coin (particles internal) dimensions, respectively~\cite{MBS02, BFM03, KMB05, WKK08}. 

Here we will define the DTQW evolution in one spatial dimension $x$ on a particle 
with $d$ distinct coin space. We will set each discrete position space step to $a$ and
discrete time step to $\delta t$ to be the same throughout the evolution of the walk.
For a $d$-dimensional system with basis state $|q\rangle$ where $q \in \{1, 2, 3, \cdots , d\}$ the spatial shift operation can be defined as
\begin{align}
 S = \sum_{q=1}^d \ket{q} \bra{q} \otimes T_{qq}.
 \end{align}
Depending on the value of $q,$ we will have two possible forms of $T_{q q}$ given by $T_-$ and $T_+$
with $\ket{x} \in \mathcal{H}_p$ (the space has to be periodic or infinite) and $\ket{q} \in \mathcal{H}_{c},$  and the coin operation
$C$ is a $ d \times d $  square unitary matrix.
As the momentum operator  $\hat{p} $  is the generator of
spatial translations in quantum mechanics we can write the components of the shift operators in the form
\begin{align}
 T_+ = e^{- i \frac{\hat{p} a}{\hbar}} = \sum_k e^{-i \frac{k a}{\hbar}} \ket{k} \bra{k}, \nonumber\\
 T_- = e^{ i \frac{\hat{p} a}{\hbar}} = \sum_k e^{i \frac{k a}{\hbar}} \ket{k} \bra{k}, \end{align} 
 where $\ket{k}$ is a momentum eigenvector with eigenvalue $ k$. 
 In the following section we will develop on this description of DTQW to simulate the neutrino oscillation probability.

 \section{ Mimicking neutrino oscillation by quantum walk}
 \label{mimic}

There are three mutually orthonormal Dirac mass eigenstates of neutrino, each of the them have two spin degrees of freedom. 
Hence, the complete neutrino dynamics in one spatial dimension is described using six internal degrees of freedom. 
To understand neutrino oscillation dynamics from DTQW perspective and simulate neutrino oscillation in any other physical system,
we need to consider system in which we can access six internal degrees of freedom. Let us define the internal space basis as 
\begin{align}
 \Big[ \ket{1,\uparrow} + 0  \ket{1, \downarrow} \Big] \oplus \Big[ 0 \ket{2,\uparrow} + 0  \ket{2, \downarrow}\Big] 
 \oplus \Big[0 \ket{3,\uparrow} + 0  \ket{3, \downarrow} \Big] \nonumber\\
 = (1 ~0 ~0~ 0~ 0~ 0~)^T  =  \ket{\zeta_1} , ~ \nonumber\\ 
 \Big[ 0 \ket{1,\uparrow} +   \ket{1, \downarrow} \Big] \oplus \Big[ 0 \ket{2,\uparrow} + 0  \ket{2, \downarrow}\Big] 
 \oplus \Big[0 \ket{3,\uparrow} + 0  \ket{3, \downarrow} \Big] \nonumber\\
 = (0 ~1 ~0~ 0~ 0~ 0~)^T =  \ket{\zeta_2}  , ~ \nonumber\\
 \Big[0 \ket{1,\uparrow} + 0  \ket{1, \downarrow} \Big] \oplus \Big[  \ket{2,\uparrow} + 0  \ket{2, \downarrow}\Big]
 \oplus \Big[0 \ket{3,\uparrow} + 0  \ket{3, \downarrow} \Big] \nonumber\\
 = (0 ~0 ~1~ 0~ 0~ 0~)^T =  \ket{\zeta_3} , ~ \nonumber\\
 \Big[0 \ket{1,\uparrow} + 0  \ket{1, \downarrow} \Big] \oplus \Big[ 0 \ket{2,\uparrow} +   \ket{2, \downarrow}\Big]
 \oplus \Big[0 \ket{3,\uparrow} + 0  \ket{3, \downarrow} \Big] \nonumber\\
 = (0 ~0 ~0~ 1~ 0~ 0~)^T =  \ket{\zeta_4} , ~ \nonumber\\
 \Big[0 \ket{1,\uparrow} + 0  \ket{1, \downarrow} \Big] \oplus \Big[ 0 \ket{2,\uparrow} + 0  \ket{2, \downarrow}\Big]
 \oplus \Big[ \ket{3,\uparrow} + 0  \ket{3, \downarrow} \Big] \nonumber\\
 = (0 ~0 ~0~ 0~ 1~ 0~)^T =  \ket{\zeta_5} , ~ \nonumber\\
 \Big[0 \ket{1,\uparrow} + 0  \ket{1, \downarrow} \Big] \oplus \Big[ 0 \ket{2,\uparrow} + 0  \ket{2, \downarrow}\Big]
 \oplus \Big[0 \ket{3,\uparrow} +   \ket{3, \downarrow} \Big] \nonumber\\
 = (0 ~0 ~0~ 0~ 0~ 1~)^T =  \ket{\zeta_6} , ~ \nonumber\\
\end{align}
where we have used the vector representation equivalence,
$$\ket{j,\uparrow} = (1 ~0)^T ~~;~~  \ket{j, \downarrow} = (0 ~1)^T~~ \text{for all}~ j = 1, 2, 3. $$
The dynamics of each flavor of the neutrino is 
defined using the Dirac Hamiltonian. Earlier results  have reported a simulation 
of two state Dirac Hamiltonian for a massive particle~\cite{cm,Marc}. 
Therefore, to simulate a six state neutrino dynamics we will form the set of three pairs,
\begin{align} 
 \text{span}\{\ket{1,\uparrow}, \ket{1, \downarrow}\}, \text{span}\{\ket{2,\uparrow}, \ket{2, \downarrow}\}, \text{span}\{\ket{3,\uparrow}, \ket{3, \downarrow}\},
\end{align}
and define a DTQW with different coin parameters for each pair.
For this purpose, we will represent the coin space in the form
\begin{align}
\mathcal{H}_{c}  = \text{span} \Big\{ \ket{\zeta_1}, \ket{\zeta_2}, \ket{\zeta_3}, \ket{\zeta_4}, \ket{\zeta_5}, \ket{\zeta_6} \Big\} =  \nonumber\\
 \text{span} \Big\{\ket{1,\uparrow}, \ket{1, \downarrow} \} \bigoplus \{\ket{2,\uparrow}, \ket{2, \downarrow}\}
 \bigoplus \{ \ket{3,\uparrow}, \ket{3, \downarrow} \Big\}. \nonumber
\end{align}
With three coins using different parameters the complete evolution operator composing of coin 
and shift operator will be in block diagonal form and one time step $(\delta t)$ of the walk operator will look like  
  \begin{align}
  \label{coinoper}
    W  = \bigoplus_{j =1,2,3} W_j =  S ( I \otimes C)  = \bigoplus_{j =1,2,3} S_j ( I \otimes C_j )  
    \end{align} where the quantum coin operation and coin state (spin) dependent position shift operators are defined as 
        \begin{align}
    C_j = \cos \theta_j \ket{j, \uparrow} \bra{j,\uparrow} + \sin \theta_j \Big ( \ket{j, \uparrow} \bra{j, \downarrow} \nonumber\\
    -  \ket{j, \downarrow} \bra{j, \uparrow} \Big) + \cos \theta_j \ket{j, \downarrow} \bra{j, \downarrow} \end{align}
    and
 \begin{align}   S_j = T_+ \otimes  \ket{j, \uparrow}\bra{j, \uparrow}  + T_- \otimes \ket{j, \downarrow}\bra{j, \downarrow}.   \end{align}

 In Eq.~(\ref{coinoper}), 
 the $j=1$ sector operates on 
 $ \text{span}\{ \ket{1, \uparrow}, \ket{1, \downarrow} \} \otimes \mathcal{H}_p,$ the $j=2$ sector operators
on $ \text{span}\{ \ket{2, \uparrow}, \ket{2, \downarrow} \} \otimes \mathcal{H}_p$ and  the $j=3$ sector operates 
on $ \text{span}\{ \ket{3, \uparrow}, \ket{3 , \downarrow} \} \otimes \mathcal{H}_p$.

The effective Hamiltonian acting on the $j$ th sector, $ H_j $ is defined as  $ \frac{\hbar}{\delta t} i \ln ( W_j ) $,
where $\hbar$ is the reduced Planck constant. $H_j$ takes the form of a 
one spatial dimensional Dirac Hamiltonian for some particular range of walk parameters~\cite{cm}. 
The coin operation is
homogeneous and the shift operator is diagonalizable in the momentum basis $\{ \ket{k}\},$ and hence 
the walk operator is diagonalizable in the same basis.

 Denoting $\frac{ k a}{\hbar} $ as $ \tilde{k},$
 the eigenvector of $H_j$ corresponding to the positive 
eigenvalue,
 $$E_j = \frac{\hbar}{\delta t} \cos^{-1} ( \cos \theta_j \cos \tilde{k} ) ~\text{for all}~j = 1 , 2 , 3  $$
 and the corresponding eigenvectors can be written as 
 \begin{align}
  \label{nustate}
  \ket{\nu_1} = & \Big( f(\theta_1, k) ~~  g(\theta_1, k) ~~0 ~~ 0 ~~ 0 ~~ 0 \Big)^T \otimes \ket{k}, \nonumber\\
  \ket{\nu_2} =  &\Big( 0 ~~ 0 ~~ f(\theta_2, k) ~~  g(\theta_2, k) ~~0 ~~ 0 \Big)^T \otimes \ket{k}, \nonumber\\
  \ket{\nu_3} = & \Big( 0 ~~ 0 ~~ 0 ~~ 0 ~~ f(\theta_3, k) ~~  g(\theta_3, k)  \Big)^T \otimes \ket{k}, 
 \end{align}
where 
\begin{align}
              f(\theta_j, k) =  \frac{\sin \theta_j e^{-i \tilde{k} }}  {\sqrt{ \sin^2 \theta_j + \Big(  \cos \theta_j \sin \tilde{k} 
              - \sqrt{ 1 - \cos^2 \theta_j \cos^2 \tilde{k}}  \Big)^2 }}  \nonumber\\
              g(\theta_j, k) =  \frac{i \Big(  \cos \theta_j \sin \tilde{k} - \sqrt{ 1 - \cos^2 \theta_j \cos^2 \tilde{k} }  \Big)} 
              {\sqrt{ \sin^2 \theta_j + \Big(  \cos \theta_j \sin \tilde{k} - \sqrt{ 1 - \cos^2 \theta_j \cos^2 \tilde{k}}  \Big)^2 }}.
\end{align}

The initial state $\ket{\Psi(0)}$ of the neutrino corresponding to electron is prepared using the operator $U$ acting on each sector,
\begin{align} \ket{\Psi(0)} = \ket{\nu_e} = \sum_{j = 1, 2, 3 }U_{e j}^{*}\ket{\nu_{j}}. 
\end{align}
This initial state is a momentum eigenstate.
After $t$ = integer $\times \delta t$ steps of the walk using the evolution operator $W$ we get
 \begin{align}
   \ket{\Psi(t)} =  W^{ \Big \lfloor \frac{t}{\delta t} \Big \rfloor } 
   \ket{\Psi(0)} = \sum_{j = 1,2,3} e^{- i \frac{1}{\hbar} E_j t} U_{ej} \ket{\nu_j}.
 \end{align}
Therefore, the survival probability of the state, $\ket{\nu_e}$ w.r.t. the time evolution is defined by 
\begin{align}\label{eoscillation}
 P_e (t) = P (\nu_e(t=0) \rightarrow \nu_e(t) ) \nonumber\\
 = |\braket{\Psi(0) | \Psi(t)}|^2 = | \sum_{j = 1,3,5} e^{- i \frac{1}{\hbar} E_j t} U_{ej} |^2.
\end{align}

Similarly, the oscillation probabilities of the other flavors are
\begin{align}\label{muoscillation}
 P_{\mu} (t) = P ( \nu_e (t=0) \rightarrow \nu_\mu(t) ) =  |\braket{ \nu_\mu | \Psi(t)}|^2 \\
  P_{\tau} (t) =  P (\nu_e(t=0) \rightarrow \nu_\tau(t) ) = |\braket{ \nu_\tau | \Psi(t)}|^2 
\end{align}
In this scheme, all the states like in Eq.~(\ref{nustate}) are evolved in the
momentum basis as it was used to implement the walk in momentum basis~\cite{Romanelli}.
In the momentum basis the shift operators are diagonal; therefore, for a state
in the momentum basis the whole walk operator will just work like a coin operator. 
 
However, if one has to implement this walk in a position basis, 
the wave function has to be distributed across positions because of the relation  
\begin{align}\label{fourier}
       \ket{k} \propto \frac{1}{\sqrt{2N + 1}} \sum_{x = -N}^{N} e^{- \frac{i k x}{\hbar}} \ket{x},
  \end{align}
 where $2N + 1$ is the total number of sites. For our description,
 position space has to be periodic or infinite. For real simulation purpose it is  
 reasonable to choose a periodic lattice with the identification $ N + 1 \equiv - N.$ 
 For that case, in place  of $x \in a ~\mathbb{Z}$ we need $x \in a ~\mathbb{Z}_{2 N + 1}$.


 From the above scheme directly we can tell that one six-dimensional quantum particle can fully simulate the neutrino oscillation mechanism.
 But experimentally it is difficult to find a six-dimensional system.
 So, we present potential many particle systems which can simulate neutrino oscillations.

{\bf Three-qubit system }
 The qubit has two degrees of freedom denoted by  $\ket{0} = ( 1 ~~ 0)^T, \ket{1} = (0 ~~ 1)^T.$   
A three-qubit system formed by tensor product of three vector space associated with each qubit will produce an eight-dimensional system.
But for simulating three-flavor neutrino oscillations we need six dimensions. So, we will confine ourselves only to the vector-space \\
 $\text{span} \Big\{ \ket{000} \equiv \ket{\zeta_1}, \ket{001} \equiv \ket{\zeta_2}, \ket{010} \equiv \ket{\zeta_3}, \ket{011} \equiv \ket{\zeta_4},
 \ket{100} \equiv \ket{\zeta_5}, \ket{101} \equiv \ket{\zeta_6}  \Big\}.$
 The coin and the shift operator which form the evolution operator of the form given in Eq.~(\ref{coinoper}) 
 for a three-qubit system can be written in the the following way : 
  \begin{align}
  C = \cos \theta_1 \ket{000}\bra{000} + \sin \theta_1 \ket{000}\bra{001} \nonumber\\
  - \sin \theta_1 \ket{001}\bra{000} + \cos \theta_1 \ket{001}\bra{001}\nonumber\\
  + \cos \theta_2 \ket{010}\bra{010} + \sin \theta_2 \ket{010}\bra{011} \nonumber\\
  - \sin \theta_2 \ket{011}\bra{010} + \cos \theta_2 \ket{011}\bra{011}\nonumber\\
  + \cos \theta_3 \ket{100}\bra{100} + \sin \theta_3 \ket{100}\bra{101} \nonumber\\
  - \sin \theta_3 \ket{101}\bra{100} + \cos \theta_3 \ket{101}\bra{101}\end{align}
and
\begin{align}
 S = T_+ \otimes \Big(  \ket{000} \bra{000} +  \ket{010} \bra{010} +   \ket{100} \bra{100}  \Big) \nonumber\\
 + T_- \otimes \Big(  \ket{001} \bra{001} +  \ket{011} \bra{011} +   \ket{101} \bra{101}  \Big). 
\end{align}
Here the coin operations $C$ and shift $S$ that act on the vector space $\text{span}\{ \ket{110}, \ket{111}\}$ are set to be zero operators. 
Thus from the complete $dim(\mathcal{H}_c) = 8$ we will be using only six dimensions. 

Therefore, the state that is equivalent to the mass eigenstates of the neutrino flavor is 
 \begin{align}
               \ket{\nu_1} =& \big( f(\theta_1, k) \ket{000} + g(\theta_1,k)  \ket{001} \big) \otimes \ket{k} \nonumber\\
               \ket{\nu_2} =  &\big( f(\theta_2, k) \ket{010} + g(\theta_2, k)  \ket{011} \big) \otimes \ket{k} \nonumber\\
                \ket{\nu_3} =  &\big( f(\theta_3, k) \ket{100} + g(\theta_3, k) \ket{101} \big) \otimes \ket{k}.     \end{align} 

{\bf Qubit$-$qutrit system }
 Similarly, we can simulate the same dynamics by a qubit$-$qutrit system. The coin space is the tensor product of coin spaces of qubit and qutrit.   
   The qubit has two degrees of freedom, $\ket{0} = ( 1 ~~ 0)^T, \ket{1} = (0 ~~ 1)^T,$ and 
  the qutrit has three degrees of freedom,  $ \ket{0} = ( 1 ~~ 0 ~~ 0 )^T,  \ket{1} = (0 ~~ 1 ~~ 0 )^T,
   \ket{2} = ( 0 ~~ 0 ~~ 1)^T,$ and together they form a six-dimensional space.
    
 For this system, $\ket{00} \equiv \ket{\zeta_1}, \ket{01} \equiv \ket{\zeta_2}, \ket{02} \equiv \ket{\zeta_3},
 \ket{10} \equiv \ket{\zeta_4}, \ket{11} \equiv \ket{\zeta_5}, \ket{12} \equiv \ket{\zeta_6}. $
 The coin and shift operator which will form the evolution operator of the form in Eq.~(\ref{coinoper}) can be written as 
 \begin{align}
 C = \cos \theta_1 \ket{00}\bra{00} + \sin \theta_1 \ket{00}\bra{01} \nonumber\\
  - \sin \theta_1 \ket{01}\bra{00} + \cos \theta_1 \ket{01}\bra{01}\nonumber\\
  +\cos \theta_2 \ket{02}\bra{02} + \sin \theta_2 \ket{02}\bra{10}  \nonumber\\
  - \sin \theta_2 \ket{10}\bra{02} + \cos \theta_2 \ket{10}\bra{10}\nonumber\\
  + \cos \theta_3 \ket{11}\bra{11} + \sin \theta_3 \ket{11}\bra{12} \nonumber\\
  - \sin \theta_3 \ket{12}\bra{11} + \cos \theta_3 \ket{12}\bra{12}
\end{align}
and
\begin{align}
 S =& T_+ \otimes \Big(  \ket{00} \bra{00} +  \ket{02} \bra{02} +   \ket{11} \bra{11}  \Big) \nonumber\\
 &+ T_- \otimes \Big(  \ket{01} \bra{01} +  \ket{10} \bra{10} +   \ket{12} \bra{12}  \Big). 
\end{align}
For this purpose, the state that is equivalent to the mass eigenstates of neutrino flavor is 
  \begin{align}
               \ket{\nu_1}& = \big( f(\theta_1, k) \ket{00} + g(\theta_1,k)  \ket{01} \big) \otimes \ket{k} \nonumber\\
               \ket{\nu_2} &=  \big( f(\theta_2, k) \ket{02} + g(\theta_2, k)  \ket{10} \big) \otimes \ket{k} \nonumber\\
                \ket{\nu_3} &=  \big( f(\theta_3, k) \ket{11} + g(\theta_3, k) \ket{12} \big) \otimes \ket{k}.
                     \end{align}

 $\sin \theta_j $s have to be small.

 \section{Numerical simulation} 
 \label{nsimulate}
 \begin{figure}
\subfigure[]{\includegraphics[width=8cm]{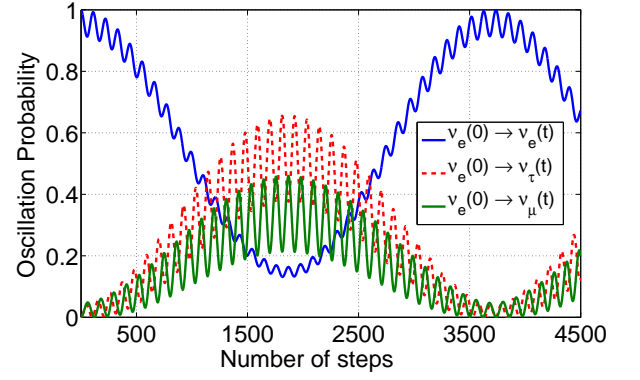}}
\subfigure[]{\includegraphics[width=8cm]{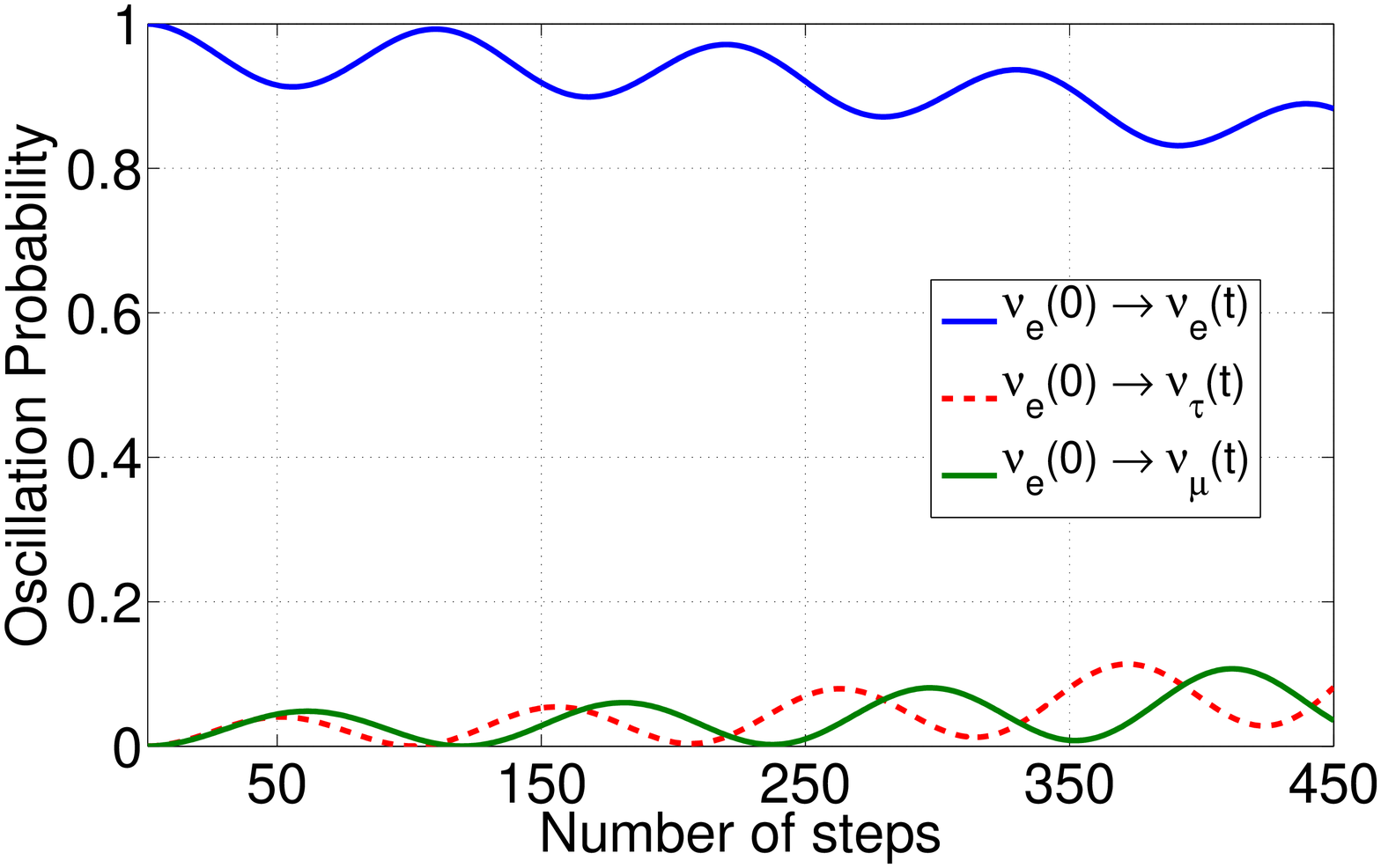}}
\caption{\small{ Oscillation probabilities obtained by numerical simulation of
DTQW for an initial state that mimics the electron neutrino. Our choices for the coin parameters to reproduce the oscillations in 
Fig. \ref{oscillation plot} are $ \theta_1 = 0.001 $ rad, $\theta_2 = 0.00615654 $ rad, $ \theta_3 = 0.0664688 $ rad. Here,
we show oscillation probability of $ \nu_e(0) \to \nu_e(t)$ {\it (blue)},
$\nu_e (0) \to \nu_\mu(t)$ {\it (green)}, $ \nu_e(0) \to \nu_\tau(t)${\it (red)}. {\bf(a)} Long range neutrino oscillation obtained for 4500
time steps of the walk.
{\bf (b)} Short range neutrino oscillation obtained for 450
time steps of the walk.}}
\label{oscillation plot from simulation}
\end{figure}

Simulation of the neutrino oscillation from DTQW can be established by finding a
correspondence \big(ideally one-to-one mapping\big) between neutrino oscillation parameters and DTQW evolution parameters.

We need to satisfy two conditions simultaneously to simulate neutrino oscillation :
(i) $ \theta_j$ and $ \tilde{k}$ should be small in Eq. (\ref{coinoper}) such that the 
DTQW produces Dirac Hamiltonian. (ii) Neutrinos are ultra-relativistic particles, 
so the relation $\tilde{k} >> \theta_j $ for all $j = 1, 2, 3$ should be satisfied.

The Dirac equation will be produced when we identify $ \theta_j = m_j c^2 ~ \frac{\delta t}{\hbar}$ and $ \tilde{k} = \frac{k a}{\hbar} = \frac{k c ~ \delta t}{\hbar}.$
Then in comparison with Eq. (\ref{phase}) of the neutrino oscillation,
\begin{align}
\Delta m_{jr}^{2}\frac{L c^3 }{4E \hbar} = \frac{( E_j - E_r )}{2} \frac{t}{\hbar} \nonumber\\
 \approx  \frac{t}{2\delta t} ~ \bigg[  \sqrt{\tilde{k}^2 + \theta^2_j } - \sqrt{\tilde{k}^2 + \theta^2_r } \bigg]    \approx
\frac{( \theta^2_j - \theta^2_r)}{4 \tilde{k}} ~ \frac{t}{\delta t}.
\end{align}  
For the case of the neutrino energy 1 GeV, $ kc = \mathcal{O}( 10^9$ eV) $ \Rightarrow    \tilde{k} = \mathcal{O} ( 10^{24} s^{-1} )~\delta t$.
Then, to have small $\tilde{k}$, $\delta t$ should be at most = $ \mathcal{O} (10^{-26} s).$
Hence, $\Delta\theta_{32}^{2} = \Delta m_{32}^{2} c^4 ~ \Big(\frac{\delta t}{\hbar}\Big)^2 \approx \mathcal{O} (10^{-25}) $,
$\Delta\theta_{21}^{2} = \Delta m_{21}^{2} c^4 ~ \Big(\frac{\delta t}{\hbar}\Big)^2 \approx \mathcal{O} (10^{-27}). $

Hence the required number of walk steps to produce a short range and a long range oscillation are 
$ \mathcal{O}( 10^{25} )$ and $ \mathcal{O}(10^{26}),$ respectively. 

For these kinds of order of $\delta t$ ,$\Delta\theta_{ij}^{2}$, 
the number of walk steps is very difficult to achieve in real lattice experiments presently. 

We should note  that, if we consider the walk time step size $\delta t = \mathcal{O}( t_p),$
the lattice space step size $ a = \mathcal{O}( l_p),$
where the Planck time = $ t_p = 5.3912 \times 10^{-44} $ s, Planck length = $ l_p = 1.6162 \times 10^{-35} $ m,
then $ \tilde{k} = \mathcal{O}(10^{-19}), \Delta\theta_{32}^2 = \mathcal{O}(10^{-59}), \Delta\theta_{21}^{2}=\mathcal{O}(10^{-61})$
and the required number of walk steps for
short and long range oscillations are  $ \mathcal{O}( 10^{42} )$ and $\mathcal{O}(10^{43}),$ respectively.
So, in principle it is possible to satisfy both
conditions (i), (ii) and simulate neutrino oscillation exactly by DTQW, but it is hard to realize in the real world.

Hopefully, the oscillation nature is determined by the quantity $\omega$t, where $\omega=\frac{E_{1}-E_{2}}{\hbar}$. 
Only the condition to simulate neutrino oscillation
is that $\omega$t will be the same in real experiment as well as in simulation system.
It implies that if we increase the frequency $\omega$, then
we can decrease the number of walk steps which can be realizable.
Thus in order to successfully simulate, we have to increase the value of the cyclic frequency, 
$$\bigg[  \sqrt{\tilde{k}^2 + \theta^2_j } - \sqrt{\tilde{k}^2 + \theta^2_r } \bigg] $$ such that 
the same oscillation profile can be obtained with a smaller number of walk steps $\frac{t}{\delta t}$.
That is to say, we are zooming in into the frequency and zooming out of the number of DTQW steps.

The Dirac dynamics is only produced by DTQW evolution when $ \theta_j$ and $ \tilde{k}$ both are small. 
 Respecting this condition, the numbers of walk steps we have chosen are 450 and 4500 for short and long range oscillation profiles, respectively.
 With the choices of parameters  $\tilde{k} = 0.01$ rad, $\theta_1 = 0.001$ rad, $\theta_2 = 0.00615654$ rad, $\theta_3 = 0.0664688$ rad. 
 In Fig.~\ref{oscillation plot from simulation} we show the neutrino oscillation probability as a function of the number of steps of DTQW.
 Both the long range and short range neutrino flavor oscillations shown in Fig.~\ref{oscillation plot} obtained from the real neutrino experiment
 and those from our DTQW simulation, Fig.~\ref{oscillation plot from simulation}, are matching perfectly. 
 
Instead of running the quantum walker for 4500 and 450 steps in a single run, we can divide the whole 
profile, respectively, in 450 and 45 runs with each run happening for 10 steps of DTQW. For that case, instead 
of taking the neutrino flavor state as the initial state for each run, 
we have to take for the $r$th run where $ r \in [1,450] $ and $ [1,45] $,
respectively, for long and short range case;
the initial state as $ W^{(r-1)10} \ket{\nu_e}.$ Else, we can store the final state,
produced at the end of $(r-1)$th run, and can start with that state, for the next run. 
We can further reduce the number of walk steps to obtain the same oscillation profile by 
going to the non-relativistic regime, where momentum can be neglected w.r.t. the masses of the neutrino \cite{MMC16a}. 
But there, the frequencies of oscillation will be proportional to the linear differences, namely $ m_j - m_l $,
not, as usual, $ m^2_j - m^2_l $.  
 
In the previous section we presented the three  possible ways of simulation using : (1) a single six-dimensional system,
(2) a three-qubit system, or (3) a qubit$-$qutrit system. All the schemes are equivalent
from the numerical simulation perspective, because all the operators are defined by  
 $dim\{\mathcal{H}_{c} \otimes \mathcal{H}_p\} \times dim\{ \mathcal{H}_{c} \otimes \mathcal{H}_p\}$
 matrices and vectors $\in \mathcal{H}_{c} \otimes \mathcal{H}_p,$ where $ dim\{\mathcal{H}_{c}\} = 6$.

\section{Entanglement entropy during neutrino oscillation}\label{entanglement}

\subsection{Entanglement between spin and position space}

In the previous sections we have assumed that all the particles are in the same momentum eigenstate $ \ket{k},$ 
but in reality they can be in a superposition of momentum eigenstates. 

For that case, we have to define the electron-neutrino state as \begin{align}\label{def}
 \ket{\nu_e} = \sum_k p(k,e) \ket{\nu_e^k} \otimes \ket{k} = \sum_{k,j} p(k,e) U^*_{ej} \ket{\nu^k_j} \otimes \ket{k}
\end{align}
where $\ket{\nu_e^k}$ denotes the spin part of electron neutrino when that is in some particular momentum eigenstate $\ket{k}.$
$\ket{\nu^k_j}$ is the spin part of the $j\text{th}$ mass eigenstate when the neutrino is in some particular momentum eigenstate $\ket{k}$.

Let us consider the initial state of the particle, 
$ \ket{\psi(0)} = \ket{\nu_e},$ then after $ \Big\lfloor \frac{t}{\delta t} \Big\rfloor$ steps of walk evolution we have the 
state 
\begin{align}
 \ket{\psi(t)} = \sum_{k,j} p(k,e) U^*_{ej} e^{- i \omega^k_j t} \ket{\nu^k_j} \otimes \ket{k}
\end{align}
where $\omega^k_j = \frac{1}{\hbar} E_j(k) $;  $ E_j(k) $ is the positive energy eigenvalue of the $j$th mass eigenstate, when the 
corresponding momentum eigenvalue $k$ is given by Eq.\,(\ref{nustate}).
Similar to the definition, Eq.\,(\ref{def}), we can define any general flavor state, 
\begin{align}
  \ket{\nu_\alpha} = \sum_k p(k,\alpha) \ket{\nu_\alpha^k} \otimes \ket{k} =  \sum_{k,j} p(k,\alpha) U^*_{\alpha j} \ket{\nu^k_j} \otimes \ket{k}.
\end{align}
The instantaneous density matrix of the system is 
 \begin{align}\label{state}
 \rho(t) = \ket{\psi(t)} \bra{\psi(t)}~~~~~~~~~~~~~~~~~~~~~~~~~~ \nonumber\\
 = \sum_{k,k',j,l} p(k,e) p^*(k',e) U^*_{ej} U_{el}  e^{- i ( \omega^k_j - \omega^{k'}_l ) t} \nonumber\\
 \Big[ \ket{\nu^k_j} \bra{\nu^{k'}_l}  \otimes \ket{k}\bra{k'} \Big].
 \end{align} 
 If we partially trace out the state with respect to the position basis (or, momentum basis),
 we have the reduced density matrix defined on $\mathcal{H}_c$,
\begin{align}\label{partialstate} 
 \rho_c(t) &= \text{Tr}_x [ \rho(t) ] =  \sum_x \braket{x | \rho(t)  | x}   \nonumber\\
 =& \sum_{k,k',j,l} p(k,e) p^*(k',e) U^*_{ej} U_{el}  e^{- i ( \omega^k_j - \omega^{k'}_l ) t} \ket{\nu^k_j} \bra{\nu^{k'}_l} \delta_{k,k'} \nonumber\\
 =& \sum_{k,j,l}  p(k,e) p^*(k,e) U^*_{ej} U_{el}  e^{- i ( \omega^k_j - \omega^k_l ) t} \ket{\nu^k_j} \bra{\nu^k_l} \nonumber\\
 =& \sum_{k,j,l} | p(k,e) |^2  W_{k}^{\Big\lfloor \frac{t}{\delta t} \Big\rfloor} \ket{\nu^k_e} \bra{\nu^k_e}  \Big( W_k^{\Big\lfloor \frac{t}{\delta t} \Big\rfloor} \Big)^\dagger,
 \end{align}
 where $ W_{k}^{\Big\lfloor \frac{t}{\delta t} \Big\rfloor} = \braket{k|   W^{\Big\lfloor \frac{t}{\delta t} \Big\rfloor}  |k} 
 =  \braket{k| W |k}^{\Big\lfloor \frac{t}{\delta t} \Big\rfloor}  .$
  
 The expression of the oscillation probability will be modified,
  \begin{align}
       \sum_k  | p(k,e) |^2 ~~ P_t(\nu_e \to \nu_\alpha, k),
   \end{align}
where $P_t(\nu_e \to \nu_\alpha, k) $ is the probability, we used in the previous sections, when neutrino selects only one momentum eigenstate.  

 The above analysis will be the same for any $\alpha$ other than $e$. 
 When $ p(k,e) = p(k,\alpha) = \delta_{k,k_0} $, 
$\rho_c(t) =  W_{k_0}^{\Big\lfloor \frac{t}{\delta t} \Big\rfloor} \ket{\nu^{k_0}_e} \bra{\nu^{k_0}_e} 
\Big( W_{k_0}^{\Big\lfloor \frac{t}{\delta t} \Big\rfloor} \Big)^\dagger,$
then the amount of entanglement between position space and internal degrees (spin-space) is always zero, as the partial traced state is pure. 
Here we will use the measure of the entanglement entropy,
\begin{align}
S_e(t) =  - \text{Tr} \Big[ \rho_c(t) ~ \log_e [ \rho_c(t) ]  \Big],
\end{align}
where $ \rho_c(t) $ is a $ 6 \times 6 $ positive semi-definite unit traced matrix.\\
So, $0 \leq   - \text{Tr} \Big[ \rho_c(t) ~ \log_6 [ \rho_c(t) ]  \Big] \leq 1$ \\
$\Rightarrow$ $ 0 \leq - \text{Tr} \Big[ \rho_c(t) ~ \log_e [ \rho_c(t) ]  \Big] \leq \log_e (6)$.
By considering a Gaussian like distribution function, the probability amplitude is defined as
\begin{align}\label{probdis}
 p(\alpha,k) = \frac{e^{-\frac{\xi}{2}(\tilde{k} - \tilde{k}_0)^2}} {\sqrt{\sum_k e^{-\xi (\tilde{k} - \tilde{k}_0)^2}}} 
\end{align}
We assumed the distribution (\ref{probdis}) 
is the same for all $\alpha = e, \mu, \tau $ as the source is the same and propagating through free space without any distortion.

\begin{figure}[h]
 \includegraphics[width = 8.5 cm]{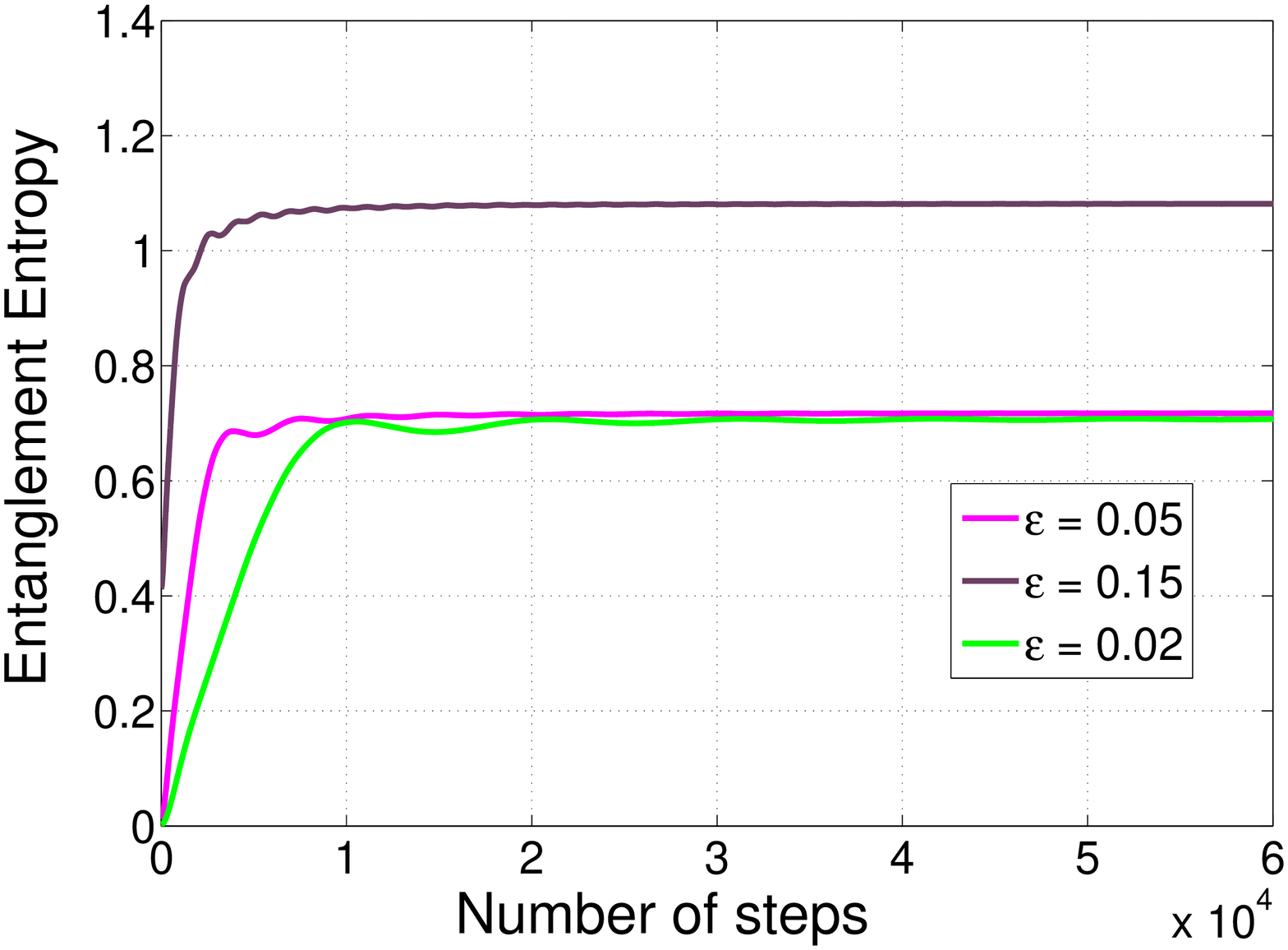}
 \caption{Entropy as a measure of the entanglement between spin and space degrees of freedom, during neutrino oscillation,
 simulated as a function of the number of DTQW steps.
 With increase in the number of steps we can see that the entanglement entropy values reache a saturation level.}
\label{ententropyfig}
\end{figure}

Our momentum eigenvalues are confined in some interval,  such that $\tilde{k} \in [ \tilde{k}_0 - \epsilon,   \tilde{k}_0 + \epsilon ]$.
$\xi$ determine the probability weight for the momentum
distribution. It is evident from the last expression of Eq.\,(\ref{partialstate}) that, increasing the value of interval,
means $\epsilon $ value will increase the corresponding 
entanglement entropy. In this sense, entanglement entropy can be used as a measure of the neutrino wave packet span in momentum space.
Larger and smaller entropy implies larger span and smaller span in momentum space, respectively. 
The wave function description in lattice space can be obtained by the Fourier transformation in momentum space, Eq.\,(\ref{fourier}), and 
the span in position space will be opposite to the span in momentum space. 
So, larger entanglement means less uncertainty in measuring instantaneous position, and a more particle-like
(localized entity) nature.

We would like to point out that the ``delocalization of the neutrino states'' discussed 
in the literature\,\cite{Kayser,Dmitry}, is related to the undetectability 
of the oscillation profile, when the oscillation wavelength (which is directly proportional to the central momentum of the wave packet,
$L^{ij}_{osc}=\frac{4\pi k_0}{\Delta m_{ij}^{2}}$) is smaller than the spread of the neutrino wave packet in position space. 
But in our case, we show the relation of entanglement among spin and space with the amount of wavepacket spreading or delocalization. 
This wavepacket spread in position space is a property of the spatial distribution 
of the neutrino source wave fuction and this is uncorrelated with the neutrino oscillation wavelength.

In Fig.\,\ref{ententropyfig} ,  we have plotted the entanglement entropy as a function of walk steps, 
for different value of parameter $\epsilon=0.02, 0.05, 0.15$  with the interval in $\tilde{k} = 0.001$.
For the numerical simulation, $\tilde{ k}_0 = 0.01$ rad, $\xi = 100$ has been used. 

From numerical simulations it is observed that for a large number of steps the measure of entanglement is almost saturating
to a fixed value and with increase in $\epsilon$ value the entanglement entropy saturates faster at higher value. 
This is a sign of the constant coupling between position space and internal degrees of freedom.
For a time varying coupling in the Hamiltonian, we can expect a deviation from saturation. 

\subsection{ Correlation between position space and particular flavor } 
\begin{figure}
\includegraphics[width=8 cm]{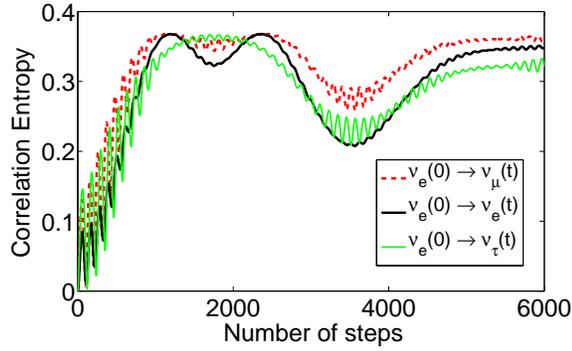}
\caption{Correlation of a particular neutrino flavor and position space as a function of the number of walk steps. 
The mean correlation values for all flavors are almost identical to one another.}
\label{flavourcorrelation1}
\end{figure}

\begin{figure}
 \includegraphics[width= 8 cm]{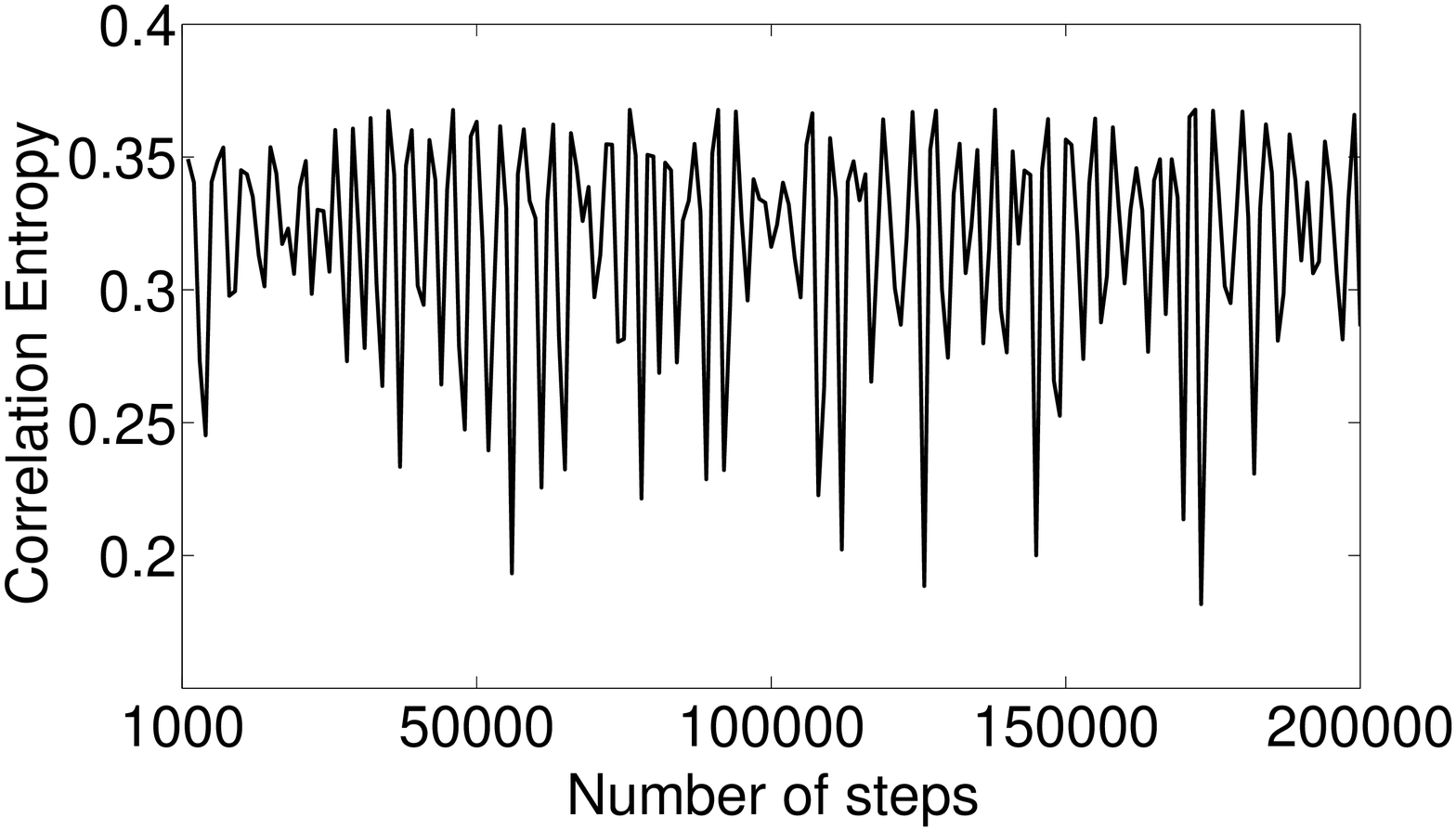}
 \caption{Correlation of a neutrino flavor $ \alpha = e$ and position space as a function of large number of walk steps. 
 Identical pattern is seen for other flavors (not shown). In a significantly large time frame we can see that the small fluctuation
 in correlation along the mean value does not follow any pattern.}
\label{flavourcorrelation2}
\end{figure}

The spin part of the $\alpha-$flavor neutrino can be defined by tracing out the momentum part;
  \begin{align}
                                      \text{Tr}_k \big[  \ket{\nu_\alpha} \bra{\nu_\alpha}  \big] 
                                      = \sum_k  | p(\alpha, k) |^2  \ket{\nu^k_\alpha}\bra{\nu^{k}_\alpha} \nonumber\\
                                      = \sum_{k,m,n}  | p(\alpha, k) |^2 ~~  U^*_{\alpha m} \ket{\nu^k_m}\bra{\nu^{k}_n} U_{\alpha n}
                                     \end{align}
  is a mixed state in general. Hence, from Eq.\,(\ref{state}), considering the projection of 
  $ \text{Tr}_k \big[  \ket{\nu_\alpha} \bra{\nu_\alpha}  \big] $,
  on the instantaneous state and tracing out the spin part 
 will give a reduced density matrix corresponding to the $\alpha$-flavor neutrino state, 
  \begin{align}\label{trace}
  \rho_\alpha(t) =  \text{Tr}_c \bigg[  \Big( \sum_k  | p(\alpha, k) |^2  \ket{\nu^k_\alpha}\bra{\nu^{k}_\alpha} \otimes \sum_{k'} \ket{k'}\bra{k'} \Big)  \rho(t)  \bigg] \nonumber\\
  = \sum_{k',k''} \bigg[ \sum_{k,m,n}   \Big\{  | p(\alpha, k) |^2  p(k',e) p^*(k'',e) \Big\} \nonumber\\
  \Big\{  U^*_{en} U_{em}  U^*_{\alpha m} U_{\alpha n} \Big\} \nonumber\\
  \braket{\nu^{k''}_m | \nu^k_m } \braket{\nu^{k}_n| \nu^{k'}_n } e^{- i ( \omega^{k'}_n - \omega^{k''}_m ) t}  \bigg]  \otimes \ket{k'}\bra{k''}.
 \end{align}
 The entropy measure, 
 \begin{align}
   S_\alpha(t) := - \text{Tr}_k \Big[  \rho_\alpha(t)  \log_e  \rho_\alpha(t) \Big] 
   \end{align}
 captures a correlation between the $\alpha$-flavor  and position-space (or momentum-space).
 In Eq.\, (\ref{trace}) we are not taking the trace over the whole coin space, we are projecting on a mixed  state 
 $\sum_k | p(\alpha, k) |^2  \ket{\nu^k_\alpha}\bra{\nu^{k}_\alpha} $, so this entropy is not actually
 the entanglement measure between $\alpha$-flavor and position space. However, we can claim that this entropy
 can still be used as a correlation measure, particularly, to comparatively understand the trend of correlations of different 
 flavor with the position space. In Fig.\,\ref{flavourcorrelation1} we show this measure of correlation
 when, $ \alpha = e , \mu, \tau$ as a function of steps of walk evolution when the value of $ \epsilon =  0.01.$ 
 In Fig.\,\ref{flavourcorrelation1} we see the increase in entropy in the beginning with increase in the number of steps and later all
 the three flavors show an identical trend in decrease and increase of the entropy around the mean value. 
 When the same measure is cosidered for a very large number of steps, shown in Fig.\,\ref{flavourcorrelation2},
 we see fluctuations around the mean value without any well-defined pattern in fluctuation 
 and these fluctuations show an identical trend for all flavors.
 From this we can say that each flavor is equally correlated with the position space during the propagation. 
 In Ref.\,\cite{Cha12} it is shown that the entanglement between the coin and position space of DTQW with strongly 
 localization but not being localized at one node (spatial disorder walk) is smaller when compared to the wide spread localized state (temporal disordered walk). 
 Therefore, from the absence of zero correlation at any point of time we can conclude that the neutrino flavor
 is not localized in position space at any given point. However, the degree of delocalization can be varied by changing the value of $\epsilon$. 
 A comparatively higher correlation would mean a more widely spread wave packet. 

\vspace{10cm}

\section{Conclusion} \label{conclu}

 Neutrinos are very weakly interacting particles, so  if the detectors are large in size detection 
 of a significant number of the neutrinos is possible.
 For neutrino oscillation experiments various kinds of detector are used; for example,
 the detector Super-Kamiokande~\cite{superkamiokande} uses 50,000 tons of ultra-pure water and  
 Sudbury neutrino observatory (SNO)~\cite{SNO} uses 1000 tonnes of ultra-pure heavy water.
 Simulating neutrino oscillation and other high energy phenomena in a low energy 
 experimental set-up gives access to intricate features of the dynamics, which is not easy in a high energy set-up.
 In this work we have shown that the three flavor neutrino oscillation obtained from this massive experimental 
 set-up can be simulated using a DTQW system with a set of walk evolution parameters. Using DTQW, 
 short range oscillations and long range oscillations have been obtained 
 by simply varying the number of steps of the walk. DTQW has been experimentally
 implemented using trapped ions \cite{tion}, cold atoms \cite{catom}, NMR \cite{nmr},
 and photons \cite{photon}; therefore, neutrino oscillations can be simulated in any of these systems.
 In addition to simulating neutrino oscillation, our work indicates that the quantum walk can play an
important role in simulating and understanding dynamics of various other physical processes in nature.
With these simulations mapping to real experimental  measurements  gives us access to exploring quantum correlations 
like entanglement and understanding the neutrino physics and high energy physics in general from the quantum information perspective.
Here we introduced a correlation measure between flavor and position space that will give information as regards the spatial degrees of 
freedom of the neutrino, by detection of a particular flavor.   
Simulating a high energy quantum dynamics in a low energy quantum system and understanding physical phenomena from
the quantum information theory perspective is an important topic of interest in contemporary research. A preprint of this paper, 
arXiv:1604.04233\,\cite{MMC16a} has already motivated research in this direction by considering the 
extension towards simulation of the neutrino oscillation in matter using DTQW \cite{MP16} without overlapping with the results shown in this paper.


 \end{document}